# Direct Probing of Polarization Charges at the Nanoscale

Owoong Kwon[1], Daehee Seol[1], Dongkyu Lee,[2] Hee Han[3], Ionela Vrejoiu[4], Woo Lee[3], Stephen Jesse,[5] Ho Nyung Lee,[2] Sergei V. Kalinin,[5] Marin Alexe[6,*], and Yunseok Kim,[1,*]

[1] School of Advanced Materials and Engineering, Sungkyunkwan University (SKKU), Suwon 16419, Republic of Korea

[2] Materials Science and Technology Division, Oak Ridge National Laboratory, Oak Ridge, Tennessee 37831, USA

[3] Korea Research Institute of Standards and Science (KRISS), Daejeon 34133 Korea

[4] University of Cologne, Köln, Germany

[5] The Center for Nanophase Materials Sciences, Oak Ridge National Laboratory, Oak Ridge, Tennessee 37831, USA

[6] Department of Physics, University of Warwick, Coventry CV4 7AL, United Kingdom

[*] Corresponding author, M.Alexe@warwick.ac.uk and yunseokkim@skku.edu

**Abstract**

Ferroelectric materials possess spontaneous polarization that can be used for multiple applications. Owing to a long term development for reducing the sizes of devices, the preparation of ferroelectric materials and devices are entering nanometer-scale regime. Accordingly, to evaluate the ferroelectricity, there is a need to investigate the polarization charge at the nanoscale. Nonetheless, it has been generally accepted that the detection of polarization charges using a conventional conductive atomic force microscopy (CAFM) without a top electrode is not feasible because the nanometer-scale radius of an atomic force microscopy (AFM) tip yields a very low signal-to-noise ratio. However, the detection is unrelated to the radius of an AFM tip, and in fact, a matter of the switched area. In this work, we demonstrate the direct probing of the polarization charge at the nanoscale using the Positive-Up-Negative-Down method based on the conventional CAFM approach without additional corrections or circuits to reduce the parasitic capacitance. We successfully probed the polarization charge densities of 73.7 and 119.0 $\mu C/cm^2$ in ferroelectric nanocapacitors and thin films, respectively. The obtained results show the feasibility of the evaluation of the polarization charge at the nanoscale and provide a new guideline for evaluating the ferroelectricity at the nanoscale.

Ferroelectric materials possess spontaneous polarization that can be applied to non-volatile memory devices owing to their electric-field induced switchable characteristics.[1-2] In addition, ferroelectric materials can be further applied to various devices because of their unique physical properties such as their piezoelectric,[3] resistive switching,[4] multiferroic,[5] and photovoltaic[6] properties. A prerequisite for these practical applications of ferroelectric materials is an evaluation of the ferroelectricity. In general, the existence of the ferroelectricity has been macroscopically examined by measuring polarization charges based on the detection of the polarization switching current.[7-9] However, owing to an important long term development for reducing the sizes of the electronic devices, the preparation of ferroelectric materials such as nanostructures and thin films are entering nanometer scale regime and accordingly, it requires the investigation of the ferroelectricity at the nanoscale level. Even though piezoresponse force microscopy (PFM) has been used extensively to explore nanoscale ferroelectricity over the past two decades,[10-12] it was recently revealed that several non-ferroelectric effects, e.g., the electrostatic effect and electrochemical strain, can also contribute to the PFM response, which often leads to a misinterpretation of the measured PFM response.[13-15] In particular, it was reported that PFM hysteresis loops can be generated even in non-ferroelectric materials.[16-17] Furthermore, new types of ferroelectric materials, e.g., $HfO_2$-based materials,[18-20] have been recently studied by many different research groups and industries to gain a fundamental understanding as well as to investigate practical applications of new ferroelectrics. Considering the current situation, an alternative approach is necessary to examine the existence of the ferroelectricity at the nanoscale level.

In this study, we demonstrate the direct probing of a polarization charge at the nanoscale level using a combination of conventional conductive atomic force microscopy (CAFM) and PFM. It has been generally accepted that the detection of polarization charge using a conventional CAFM set-up without a top electrode is not feasible because the nanometer-scale radius of an AFM tip leads to a very low signal-to-noise ratio. However, in ferroelectric materials, the detection of the polarization charge may not be related to the radius of an AFM tip, and it is instead an important issue related to the switched area. To prove this concept, we first choose ferroelectric $BiFeO_3$ (BFO) nanocapacitors with a radius of about 190 nm because they can generate relatively large switching currents owing to the relatively larger area of the top electrode compared to that of the AFM tip. We successfully obtained a remnant polarization charge density of 73.7 $\mu C/cm^2$ for the ferroelectric nanocapacitors by using the Positive-Up-Negative-Down (PUND) method[21-25] based on the conventional CAFM approach, which we refer to as AFM-PUND. Furthermore, by performing frequency-dependent measurements, we confirmed that the measured switching current was the real switching current, and we also found that the switching current increases with the switched area, even beyond the top electrode area. Eventually, using the proposed AFM-PUND approach, we successfully probed the polarization charge directly, of which remnant density is 119.0 $\mu C/cm^2$, in ferroelectric $Pb(Zr_{0.2}Ti_{0.8})O_3$ (PZT) thin films. These results can provide a new guideline for evaluating the polarization charge at the nanoscale level based on the AFM approaches.

To demonstrate the feasibility of nanoscale probing of polarization charge based on the AFM, we chose ferroelectric BFO nanocapacitors as a first model system because the switched area is still on the submicron-scale range; however, it can be increased up to the top electrode area, which ranges from 0.09–0.12 $\mu m^2$ in this study. In other words, owing to the relatively large switched area when compared to the contact area of the AFM tip, the probability of detecting the switching current can be much higher in this sample.

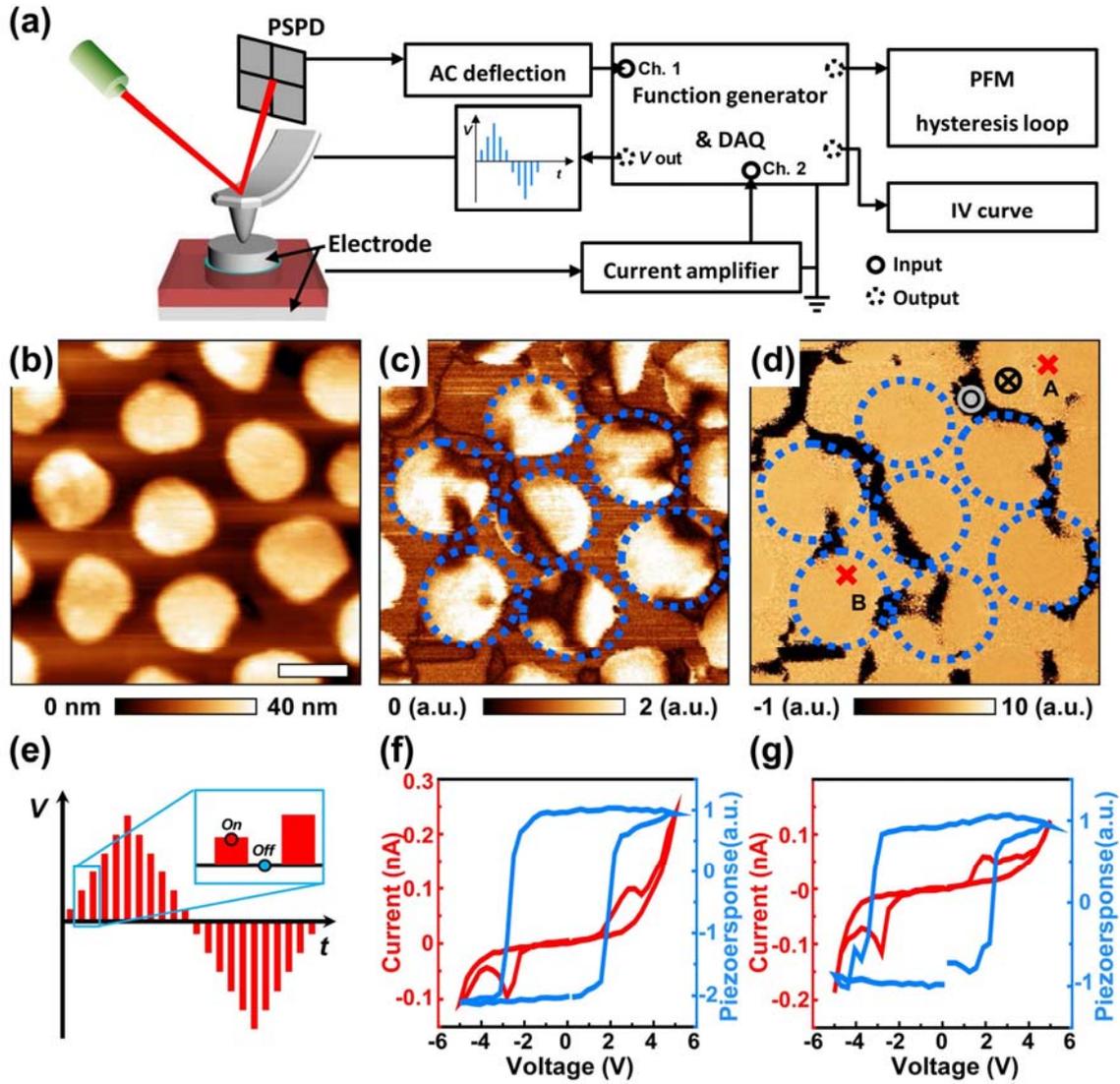

**Figure 1.** (a) Schematic of PFM hysteresis loop and I-V curve measurements. (b) Topography,

(c) PFM amplitude, and (d) PFM phase images of the as-prepared ferroelectric nanocapacitors.

(e) Applied triangular waveform for acquired PFM hysteresis loops and I-V curves. (f, g) I-V

curves and PFM hysteresis loops measured from points (f) A and (g) B in Figure 1(d),

respectively. The piezoresponse is obtained from the product of the amplitude and phase. Scale bar in Figure (b) is 320 nm.

Prior to measuring the polarization charge, we evaluated the ferroelectric properties and I-V characteristics of the prepared sample through a conventional AFM equipped with a function generator and a data acquisition (DAQ) system, as shown in Figure 1(a). We applied the bipolar triangular waveform (Figure 1(e)) to the sample using the conductive AFM tip, and we then simultaneously acquired the PFM response and current using the DAQ. Figures 1(b-d) show the topography and corresponding PFM amplitude and phase images of the as-prepared ferroelectric nanocapacitor arrays. The topography image clearly shows that there are nanocapacitor arrays with radii of around 190 nm and height of around 20 nm.[26] We observed multidomain structures mixed with up- and downward directions within the nanocapacitors in the PFM amplitude and phase images. As shown in Figures 1(f, g), S1, S2, and S3, the ferroelectric switching characteristics were locally different, and were dependent on the as-prepared domain structures of the nanocapacitors.[26-27] I-V characteristics were also locally varying, as shown in Figures 1(f, g) and S4. Interestingly, in most of the I-V curves, we were able to clearly see current peaks, as shown in Figures 1(f, g). It is believed that the observed current peak corresponds to the coercive voltage that can be deduced from the PFM hysteresis loops (see Figures 1(f, g)). From a comparison of spatial images of the coercive (Figure S3) and current peak voltages (Figure S4), we confirmed that the current peak corresponds to the coercive voltage (Figure S5), indicating that the current peaks in Figures 1(f, g) can be regarded as switching current peaks.

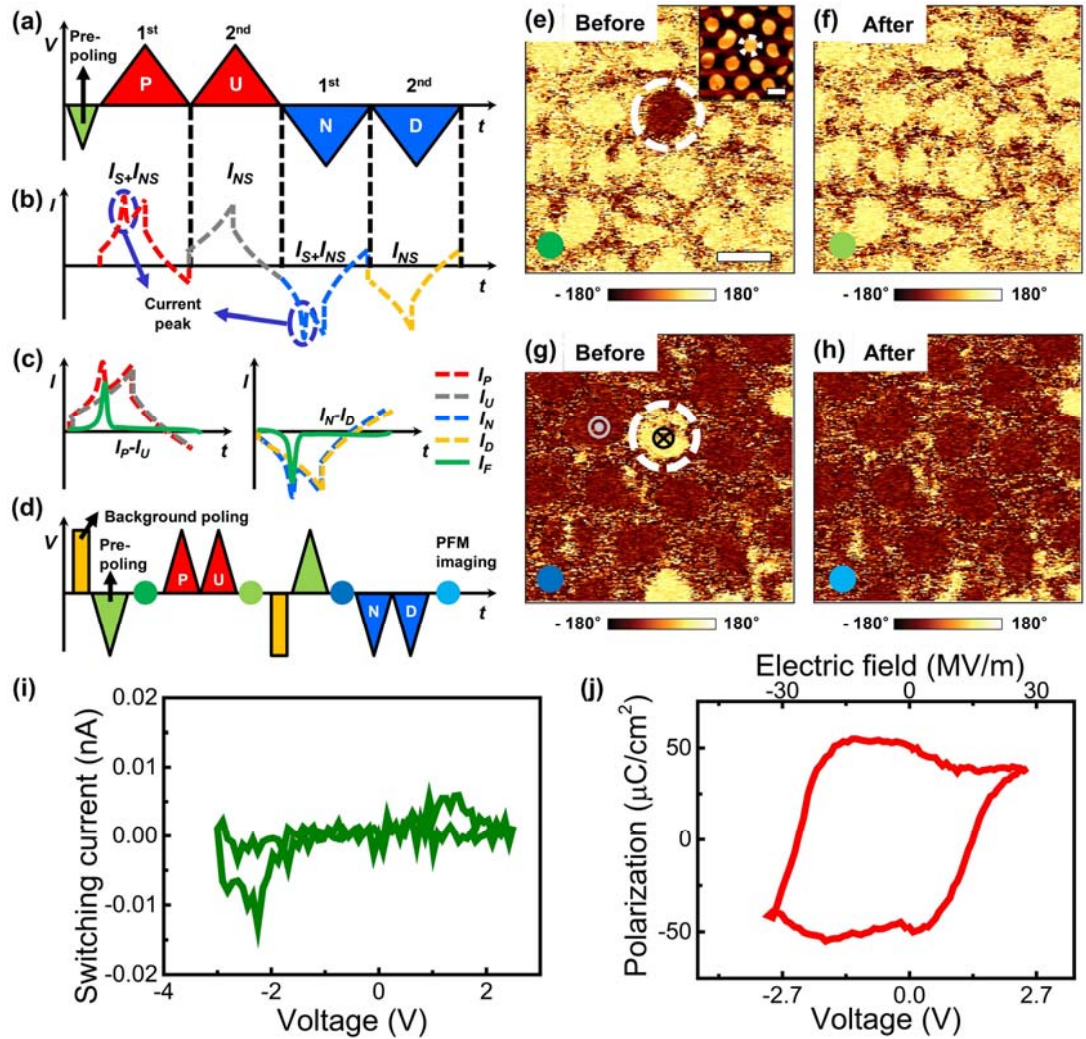

**Figure 2.** Investigation of ferroelectric nanocapacitors: (a-c) Schematics of (a) a typical PUND waveform for macroscopic P-E hysteresis loop measurement, (b) current acquired using PUND, and (c) switching current obtained by subtracting non-switching currents. (d) Schematic of AFM-PUND sequence. The colored circles in Figures (d) and (e-h) indicate PFM imaging stages. (e-h) PFM phase images (e, f) before and (g, h) after applying (e, g) *PU* and (f, h) *ND* waveforms in the ferroelectric nanocapacitors. The white dotted circles in Figures. (e) and (f) represent the locations of the target ferroelectric nanocapacitor for the measurement. Inset in Figure (e) shows the topography of the measured region. (i)

Ferroelectric switching current acquired by the AFM-PUND waveform and (j) corresponding P-E hysteresis loop. Prior to the AFM-PUND measurement, to clearly show and confine the switched area within one nanocapacitor, background poling was first performed by applying voltages of (e) 3.5 V and (f) -3.5 V, respectively, over an entire area, and then pre-poling was conducted by applying a unipolar triangular waveform with a minimum of (e) -3 V and a maximum of (f) 2.5 V, respectively, to the single nanocapacitor, as seen in Figures. (e) and (f). However, note that because the aim of the background poling is to clearly observe the switched area in the PFM phase images, it is not necessarily required prior to the AFM-PUND measurement. The scale bar is 400 nm.

In order to investigate the polarization charge based on the observed switching current (Figures 1(f,g)), we employed the PUND method.[24, 28] In conventional polarization-electric field (P-E) measurements using a Sawyer-Tower circuit, the contribution of non-switching currents cannot be fully eliminated during the polarization-charge measurement. However, with the PUND method, because non-switching contributions can be entirely eliminated, the polarization charge can be explicitly separated from the total measured current. Specifically, to separate the contribution of the switching current, a triangular unipolar waveform with the same polarity is consecutively applied two times on the sample (Figure 2(a)). This procedure can be explained using the following equations[24]:

$$I_1 = I_S + I_{NS} \tag{1}$$

$$I_2 = I_{NS} \tag{2}$$

where $I_1$ and $I_2$ are the total measured current from the 1$^{st}$ (positive ($P$) and negative ($N$)) and 2$^{nd}$ (up ($U$) and down ($D$)) waveforms, respectively, and $I_S$ and $I_{NS}$ are current components that are induced by ferroelectric polarization switching and non-switching contributions, including dielectric and and leakage currents, respectively.[20,24,29,30] When the 1st $P$ unipolar waveform is applied to the sample with upward polarization, it can induce downward polarization switching. In this step, the measured current includes both switching and non-switching contributions, as shown in Eq. (1). In general, the polarization-switching charge appears as a current peak. In the 2nd $U$ unipolar waveform, the current of non-switching components are acquired because the polarization is already switched in the 1st $P$ step (see Figure 2(b)).    Therefore, by subtracting $I_1$ and $I_2$, the contributions of the non-switching current are removed (see Figure 2(c)). Then, the polarization switching charge can be readily calculated from the separated switching current using the equation below:

$$\Delta Q = \Delta I \cdot t \qquad\qquad (3)$$

where $Q$ and $t$ are the respective charges and time of the applied voltage (in the present case, it corresponds to a single voltage step for 1 ms). Based on the above procedures, polarization charges can be obtained, and a corresponding P-E hysteresis loop can be constructed.

Based on the PUND method, the AFM-PUND sequence was performed as shown in Figure 2(d). Prior to the application of triangular waveforms for the AFM-PUND measurements, we performed background poling by applying a positive (negative) bias over the entire area of the nanocapacitor array. Then, we applied a negative (positive) unipolar waveform for the pre-poling procedure in order to switch the as-grown polarization to the upward (downward) direction at the target nanocapacitor. Finally, we acquired the switching current by applying the 1st ($P$ and $N$) and 2nd ($U$ and $D$) waveforms. It should be noted that background poling is not necessarily required for the PUND measurement in some cases. While the background poling was performed over the entire measured area to enable a clear observation of the switched area, the pre-poling was performed solely in the target nanocapacitor.

After subtracting the current as described above, we were able to separate the switching current component, as shown in Figure 2(i). In the switching current, a current peak was visible in each side of the applied voltage. To convert the current into a polarization charge, an evaluation of the actual switched area is necessary. In our case, because we performed pre-poling for the target nanocapacitor, the actual switched area may correspond to the top electrode area of the nanocapacitor. However, this fact was double-checked by PFM imaging before and after application of the *PU* and *ND* waveforms to the nanocapacitor. As shown in Figures 2(e-g), the switched area was confined within one nanocapacitor by pre-poling for both polarities, and the actual switched areas were around 0.113–0.121 $\mu m^2$, which corresponds to the top electrode area of the nanocapacitor. (See Figure S6) Based on the measured switching current and PFM phase images, we could obtain the polarization charge and corresponding P-E hysteresis loop (Figure 2(j)). The remnant polarization charge densities for the nanocapacitor in Figure 2 were around 55 $\mu C/cm^2$ and the average value for fifteen nanocapacitors is 73.7±11.1 $\mu C/cm^2$. We note that the deviation of the average value is relatively high because each nanocapacitor shows very different switching characteristics (see Figures S3, S4, and S5). Nonetheless, the remnant polarization charge densities are within the reasonable range of 50–80 $\mu C/cm^2$ obtained by the macroscopic P-E measurement.[31-33] Note that the investigation of the polarization charge as well as the P-E hysteresis loop were performed by conventional CAFM and PFM set-ups without additional corrections or circuits to reduce parasitic capacitance.[32,34,35]

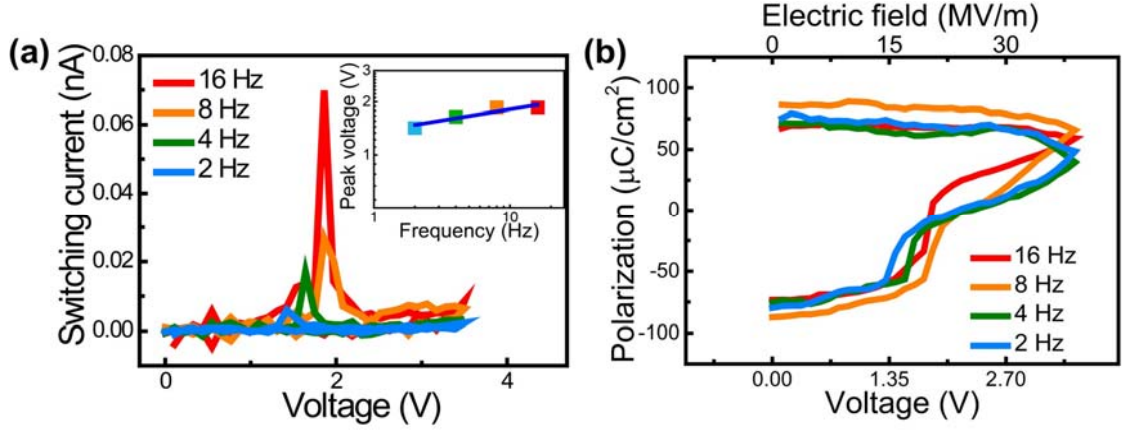

**Figure 3.** (a) Frequency-dependent polarization switching current acquired by the AFM-PUND waveform and (b) corresponding P-E hysteresis loops in the ferroelectric nanocapacitors. Inset in Figure 3(a) shows the current peak voltage, *i.e.,* coercive voltage, versus frequency of the applied AFM-PUND waveform. Note that the measured nanocapacitor is different from that in Figure 2.

We further measured the frequency-dependent switching current to reconfirm that the current peaks in Figure 2(i) originate from the polarization switching. If the AFM-PUND method is correctly performed, the ferroelectric switching current should increase with the increasing frequency of the waveform.[36] As expected, the switching current increases with increasing frequency, *i.e.,* decreasing pulse width, as presented in Figures 3(a) and S7. Furthermore, the switching current peak voltage, *i.e.,* the coercive voltage, follows the frequency dependence as shown in the following equation[37]:

$$\ln\left(V_{peak}\right) = A \cdot \ln(f) + B \tag{4}$$

where $V_{peak}$ and $f$ are the switching peak voltage and frequency of the waveform, respectively, and $A$ and $B$ are the exponent coefficient and offset, respectively. The fitting of the peak voltage using Eq. (4) in the *ln(V_{peak})* versus *ln(f)* plot shows that the exponential coefficient and offset are 0.12688 and 0.30327, respectively (inset of Figure 3(a)). In the previous reports[36], the value of the exponential coefficient in the ferroelectric material is around 0.05–0.23. This indicates that our results are in good agreement with the previous reports[38-41]. In addition, even though the constructed P-E hysteresis loops show slightly different shapes, the remnant polarization values are within a similar range, i.e., around 70 μC/cm² for this target nanocapacitor. As a result, we confirm that the polarization charge can be investigated using conventional CAFM and PFM set-ups at the nanoscale level.

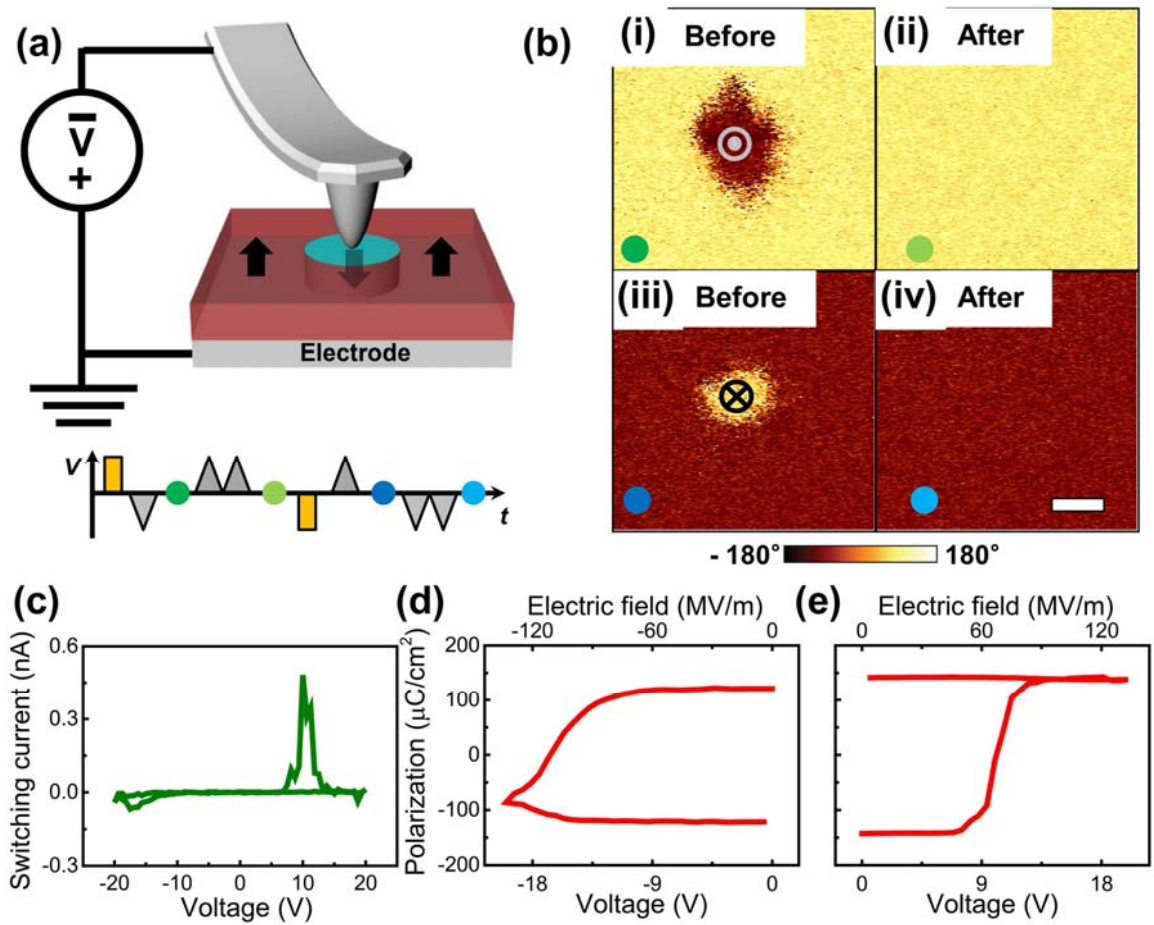

**Figure 4.** (a) Schematics of (bottom) the AFM-PUND waveform and (top) AFM-PUND measurement in the ferroelectric thin film. (b) PFM phase images (i, ii) before and (iii, iv) after applying (i, iii) *PU* and (ii, iv) *ND* waveforms and a corresponding (c) switching current in the ferroelectric thin film. The colored circles in Figures (a, b) represent PFM imaging stages in the AFM-PUND sequence. (d,e) P-E hysteresis loop constructed from the switching current of (d) negative and (e) positive bias branches, respectively, in Figure (c). Note that PFM images of the switched area were acquired using the Pt/Cr coated tip before and after the switching current measurement through the pure Pt tip. Similar to the case of the nanocapacitor, the background poling was first performed by applying (b (i)) 20 V and (b (ii)) -20 V, respectively, over an entire area, and then the pre-poling was conducted by applying a unipolar triangular waveform down to a minimum of (d) -20 V and up to a maximum of (e) +20 V, respectively, prior to the AFM-PUND measurement. The scale bar is 400 nm.

Finally, we investigated also the polarization charges in the ferroelectric PZT thin film using the AFM-PUND method. In the capacitor structures, the fully switched area is regarded as the top electrode area of the capacitor, indicating that the switched area is determined by the top electrode area. In a similar manner, the switched area for the thin film case can be primarily determined by the contact area of the AFM tip because the AFM tip acts as a top electrode. Because the minimum switched area is typically determined by the radius of the AFM tip, the use of an AFM tip with a relatively large radius can further increase the switched area. Note that the reason for which we were able to measure the polarization charge in the nanocapacitor is that the switched area is relatively large, enabling us to measure the switching current even if the nanocapacitor area is still within the submicron-scale range.

Thus, to physically increase the contact area and the corresponding minimum switched area, we used a blunted pure Pt tip with a radius of 60 nm (Figure S8) for the AFM-PUND measurement in the thin film. Note that because the tip is a pure Pt tip and its shape is normal, we did not expect there to be any serious electrical contact issue.

Furthermore, there may be another issue in the AFM-PUND measurement. As similar for the nanocapacitors shown in Figures 2 and 3, both background and pre-poling procedures were performed. In particular, we performed the pre-poling procedure in the PZT thin film although ferroelectric materials show a preferred direction. This is necessary to predetermine the back-switched area during the sequence of the *P* and *N* waveforms as well as avoid further switching during the sequence of the *U* and *D* waveforms. If there is no pre-poling procedure, the switched area will continuously increase under the repeated runs of the applied waveforms. For instance, the switching current continues to be detected as we increase the number of applications of the *U* and *D* waveforms because the switched events keep happening (see the details in Figure S9). Thus, in order to limit the switched area of the PZT thin film, we performed the pre-poling procedure for the thin film with a preferred polarization direction. Then, we applied *PU* and *ND* waveforms and measured the current at each target area using a pure Pt tip (see Figure S11). Before and after the application of the AFM-PUND waveforms, PFM images were obtained to evaluate the switched area during the pre-poling procedure and its back-switching to the background state during the *PU* and *ND* waveforms. As shown in Figure 4(b), the switched area is much larger than the area of the pure Pt tip. The corresponding switching current peaks are clearly visible, even in the ferroelectric thin films, as shown in Figure 4(c). A relatively large peak was generated in the positive bias branch because the switched area of the positive bias branch ($0.583 \ \mu m^2$) is larger than that of the negative bias branch (0.324 $\mu m^2$). This asymmetric current might be also relevant to the asymmetric junction properties of top (i.e., AFM tip) and bottom electrodes with the ferroelectric thin film.

Then, the question is whether the obtained switching current can be properly used to evaluate the polarization charge when the switched area is much larger than the area of the top electrode, i.e., the contact area of the AFM tip. In many cases, the switched area may be much larger than the contact area of the AFM tip, and it can be readily modulated by the environmental conditions, *e.g.,* humidity[42-43] and/or applied voltage conditions[44-45]. As reported in an earlier report[44] and Figure S12, the switched area, even in the capacitor structures, may be further extended toward the outside of the capacitor boundary under a relatively higher or longer voltage conditions. For such a case, if the actual switched area is obtained from the PFM image, the calculated polarization charge density is within the acceptable range (see the details in Figure S13). Therefore, although the switched area is much larger than the contact area of the AFM tip in the thin film, the polarization charge density calculated from the switching current is within the reasonable range. Considering the switched area of the ferroelectric thin film in Figure 4, the polarization charge density was calculated to be around 120 $\mu C/cm^2$. The average value for fifteen times measurements obtained through *PU* waveform is 119.0±8.3 $\mu C/cm^2$ for the positive bias branch, which is close to the previously reported values (albeit, slightly higher),[46-47] as shown in Figures 4(d,e). Because we obtained the switched area from the PFM phase images, this area value may be slightly underestimated. We note that the relative deviation compared to the average value for the ferroelectric thin film (7.0 %) is much lower than that for the ferroelectric nanocapacitor (15.1 %) because of the locally different switching properties in the BFO nanocapacitors. The accurate calibration for the proposed AFM-PUND approach should be further studied for obtaining accurate polarization value. We also note that slightly asymmetric polarization charge densities might originate from the asymmetric current in Figures 4(d,e) related to the asymmetric junction properties of top (i.e., AFM tip) and bottom electrodes with the ferroelectric thin film. Nevertheless, we successfully measured the polarization charge in the ferroelectric thin film at the nanoscale level. Consequently, these results could provide new insight for the exploration of ferroelectricity at the nanoscale level, and may contribute to the resolution of complicated issues pertaining to the interpretation of the PFM hysteresis loop.

In conclusion, in this study, we clearly showed that direct probing of the polarization charge is possible at the nanoscale level using the proposed AFM-PUND method. In the ferroelectric nanocapacitor sample, we observed current peaks in the I-V curves and verified that they originated from the ferroelectric polarization switching. Then, based on the measured switching current and the switched area, using the combination of the proposed AFM-PUND method and the conventional PFM, we successfully calculated the polarization charge density, which is within the reasonable range when compared to the macroscopic one in the nanocapacitor. Finally, we showed that the polarization charge can also be measured even in the ferroelectric thin film by confining and increasing the switched area using pre-poling and the pure Pt tip with a relatively large radius, respectively. To the best of our knowledge, our finding is the first to show that nanoscale-level polarization charge measurements are possible using conventional CAFM and PFM set-ups. Consequently, it is a very powerful and simple way of studying the presence of ferroelectricity at the nanoscale level.

**Experimental Section**

Ferroelectric nanocapacitor: An epitaxial BFO thin film with a thickness of 90 nm was grown on a 75-nm thick $SrRuO_3$ (SRO) bottom electrode deposited on a (001) oriented $SrTiO_3$ (STO) substrate by pulsed laser deposition (PLD). Ultrathin anodic aluminum oxide (AAO) shadow masks were prepared by anodization of aluminum and placed on the BFO thin film. Then, 25-nm thick Pt top electrodes were deposited through the AAO shadow mask by electron-beam evaporation. Finally, film-type Pt/BFO/SRO nanocapacitors with a radius of around 190 nm were obtained by removing the AAO mask. Details of the fabrication conditions can be found elsewhere[26, 48].

Ferroelectric thin film: An epitaxial PZT film (150 nm in thickness) was grown on a single crystal of (001) 0.5% Nb-doped STO conducting substrate using PLD. Growth of the film was conducted at 625°C in an oxygen partial pressure of 100 mTorr. The laser fluence and repetition rate were fixed at 1.0 $J/cm^2$ and 5 Hz, respectively. After completing the PZT deposition, the sample was cooled to room temperature in the PLD chamber for ~1 h under an oxygen partial pressure of 100 Torr. Details on the thin film growth can be found elsewhere.[47]

AFM measurements: AFM measurements were carried out using commercial systems (Asylum Research Cypher and Park Systems NX10) equipped with a LabVIEW/MATLAB-based band excitation (BE) controller and a current amplifier (Femto DLPCA-200). In both systems, a conductive Pt/Cr-coated AFM probe (BudgetSensors, Multi75E-G) with a spring constant of 3 N/m was used. In the case of the nanocapacitors, PFM measurements were conducted by applying an ac modulation bias of 0.4 V with 340 kHz (Figure 1). In Figure 2, PFM measurements were performed with a lock-in amplifier (Stanford Research Systems SR-

830) and applying an ac moculation bias of 0.5 V with 17 kHz. BE piezoresponse force spectroscopy (BEPS) measurements were carried out by applying bipolar triangular waveforms up to ±5 V with a BE waveform of 0.4 V at 340‑420 kHz. To clearly observe the polarization switching behavior, two hysteresis loops were successively acquired at each individual grid point, and the average of the second loop was then displayed. I-V curves were simultaneously measured with BEPS measurements. In the case of the films, PFM images were taken upon the application of an ac modulation bias of 1.5 V at 17 kHz.

Prior to the AFM-PUND measurements, background poling was first performed over an entire area to align each polarity, and then pre-poling was conducted by applying a unipolar triangular waveform to align the opposite polarity. Then, AFM-PUND measurements were obtained by applying unipolar positive and negative waveforms, which are composed of 64 steps with a pulse duration of 1 ms per step (~16 Hz) and are similar to that in Figure 1(e). Note that a pure Pt tip (Rocky Mountain Nanotechnology 25Pt400B) was used for the AFM-PUND measurement in the PZT thin film. A pristine pure Pt tip was worn out by performing 25 consecutive scans with a set-point of 700 nN. For the frequency-dependent *PU* measurements, unipolar triangular waveforms composed of 64 steps with different pulse durations (1, 2, 4, and 8 ms) were applied to the nanocapacitors.

## Acknowledgements

This work was supported by the Basic Research Lab. Program through the National Research Foundation of Korea (NRF) funded by the Ministry of Science, ICT & Future Planning (NRF-2014R1A4A1008474). A part of this research was supported by the U.S. Department of Energy, Office of Science, Basic Energy Sciences, Materials Sciences and Engineering

Division (ferroelectric film preparation) and was conducted at the Center for Nanophase Materials Sciences, which is sponsored at Oak Ridge National Laboratory by the Office of Basic Energy Sciences, U.S. Department of Energy. M.A. acknowledges the financial support of the Royal Society through Wolfson Award and Theo Murphy award.

**Supplementary materials**

# Direct Probing of Polarization Charge at the Nanoscale

Owoong Kwon[1], Daehee Seol[1], Dongkyu Lee,[2] Hee Han[3], Ionela Lindfors-Vrejoiu[4], Woo Lee[3], Stephen Jesse,[5] Ho Nyung Lee,[2] Sergei V. Kalinin,[5] Marin Alexe[6,*], and Yunseok Kim,[1,†]

[1] School of Advanced Materials and Engineering, Sungkyunkwan University (SKKU), Suwon 16419, Republic of Korea

[2] Materials Science and Technology Division, Oak Ridge National Laboratory, Oak Ridge, Tennessee 37831, USA

[3] Korea Research Institute of Standards and Science (KRISS), Daejeon 34133 Korea

[4] University of Cologne, Köln, Germany

[5] The Center for Nanophase Materials Sciences, Oak Ridge National Laboratory, Oak Ridge, Tennessee 37831, USA

[6] Department of Physics, University of Warwick, Coventry CV4 7AL, United Kingdom

---

[†] Corresponding author, M.Alexe@warwick.ac.uk and yunseokkim@skku.edu

**1. Local switching behavior of ferroelectric nanocapacitors**

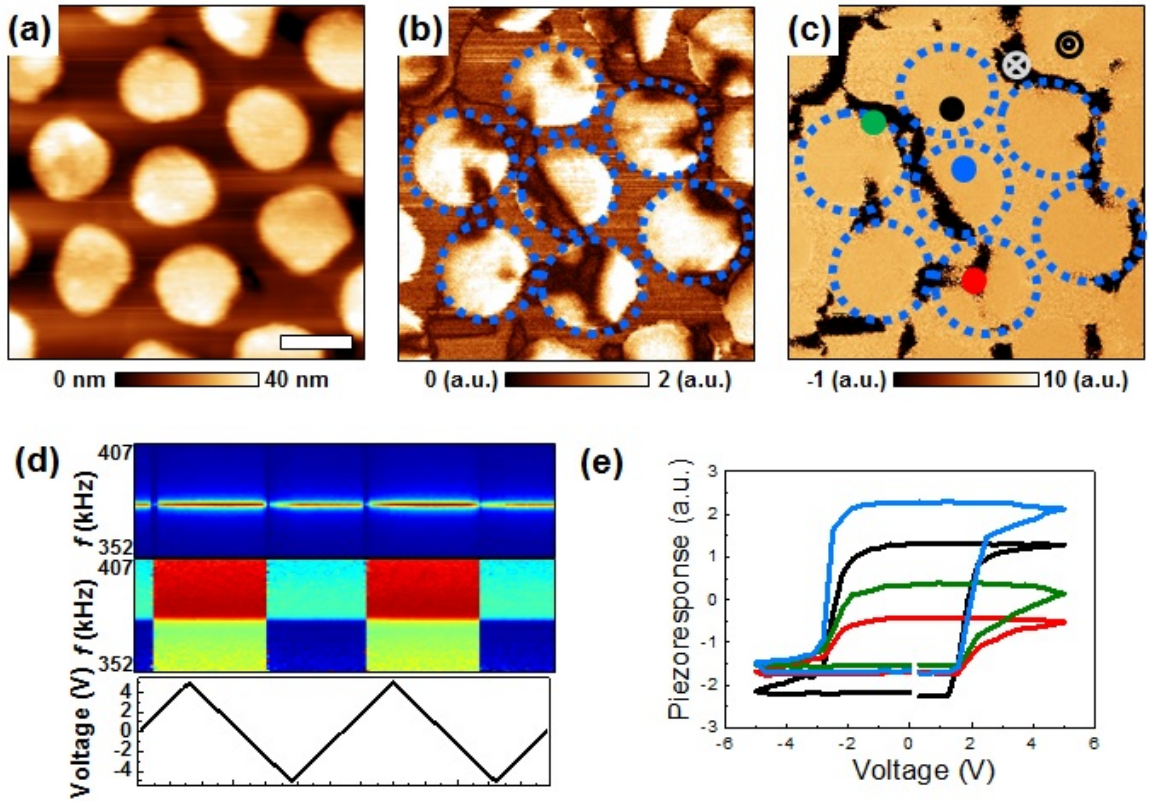

**Figure S1**. (a) Topography, (b) PFM amplitude, and (c) PFM phase images of the as-prepared nanocapacitors. The scale bar represents 320 nm. (d) BE amplitude (top), phase (middle), and applied triangular waveform (bottom). The spectra were obtained from the blue circle of Figure S1(c). (e) PFM hysteresis loops obtained from each circle of Figure S1(c). The dotted lines represent the locations of the nanocapacitors.

Figures S1(b) and S1(c) show the PFM amplitude and phase images of the nanocapacitor. Note that Figures S1(a-c) are the same images as Figs. 1(b-d). The amplitude and phase spectrum in Figure S1(d) prove that there is clear ferroelectric polarization switching, which is indicated by the straight vertical lines of the amplitude and the phase difference of 180°, respectively[1]. In the PFM hysteresis loop in Figure S1(e), even though each of the nanocapacitors shows locally different switching properties, the clear shape of PFM hysteresis loops was observed for each case based on the polarization switching.

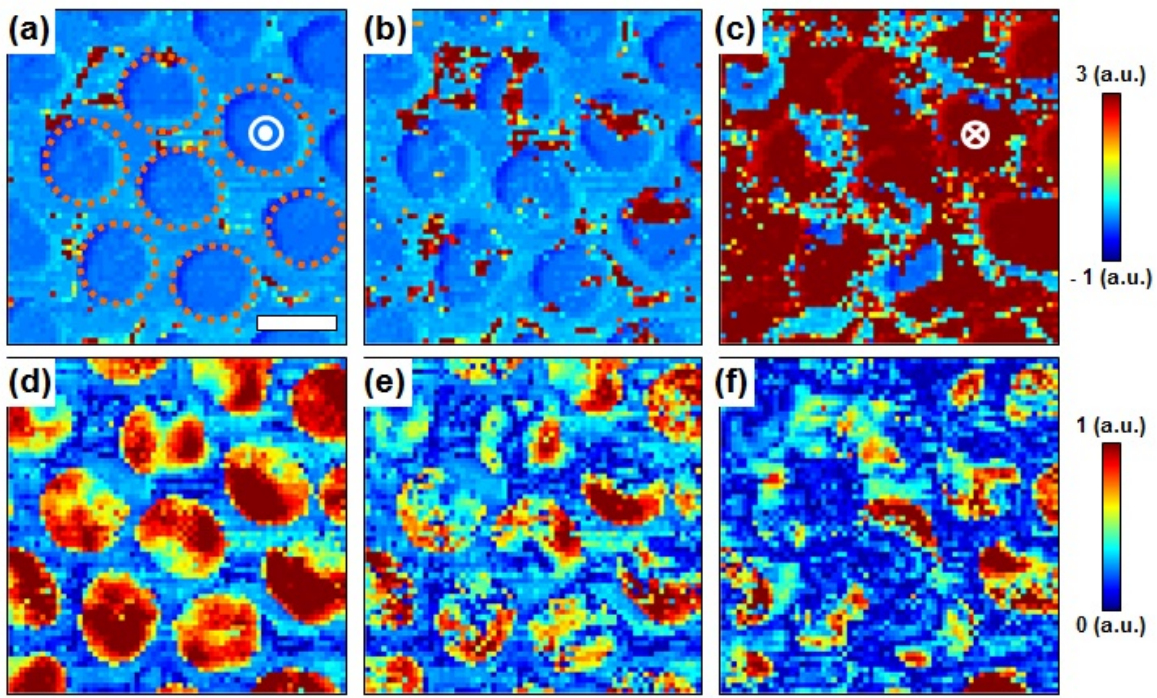

**Figure S2**. Snapshot images of BEPS (a-c) phase and (d-f) amplitude at (a,d) 1.250 V, (b,e) 1.876 V, and (c,f) 4.0625 V applied to the conductive AFM tip in the ferroelectric nannocapacitors. The orange dotted lines in Figure (a) represent the locations of the nanocapacitors arrays. The scale bar represents 320 nm.

Figure S2 shows selected snapshot images of the BEPS phase and amplitude at each voltage. The BEPS data were acquired on 64 × 64 grid points over an area of 2 μm × 2 μm. As shown in Figure S2(a), most of the area shows an upward polarization under an applied dc voltage below the coercive voltage, in this case, it is +1.250 V. In addition, the amplitude of Figure S2(d) shows a strong response within the nanocapacitors. This indicates that there is no significant switching over the measured area under the applied dc voltage below the coercive voltage. However, as we increase the applied dc voltage above the coercive voltage, some parts of the area begin to be back switched (see red contrast in Figure S2(b)). Under the applied dc voltage of around +4.0625 V, most of the area was switched as opposite polarizations (see Figure S2(c)). Accordingly, the amplitude becomes weaker than in the as-prepared state, as shown in Figures S2(d) and S2(f). These observations indicate that even though the polarization switching is locally varied, the applied dc voltage above the coercive voltage induces the polarization switching in most of the measured area.

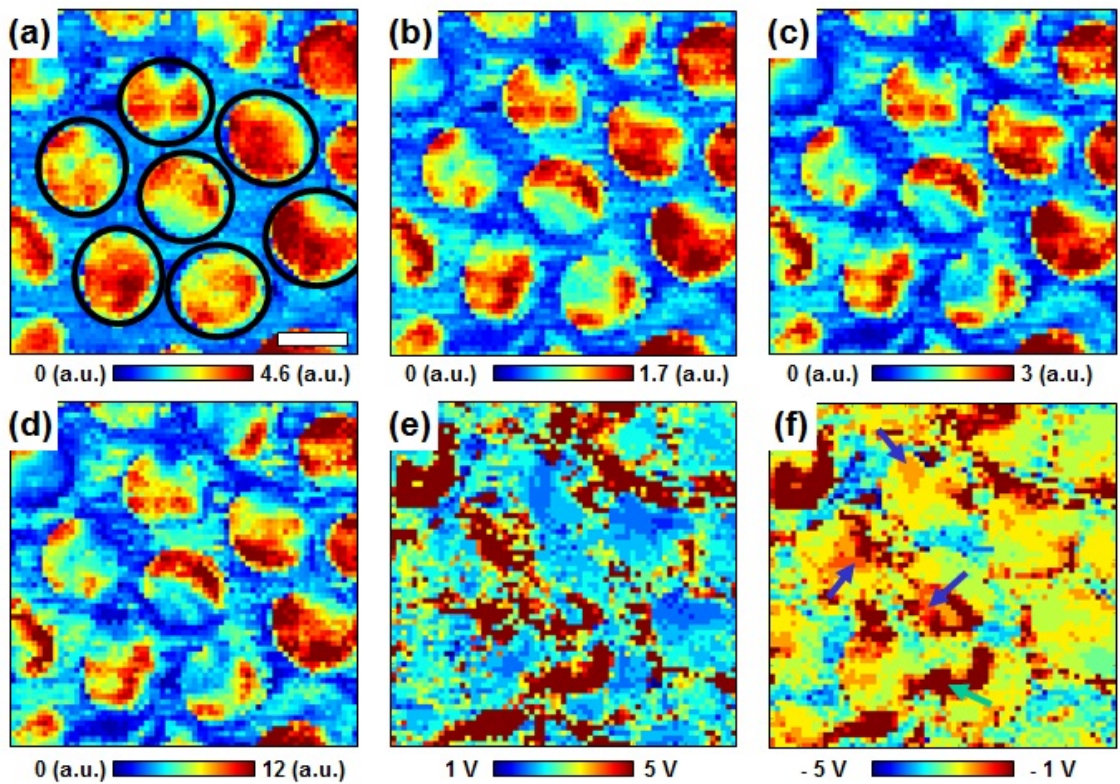

**Figure S3**. BEPS spatial maps of the nanocapacitors: (a) positive remnant piezoresponse, (b) negative remnant piezoresponse, (c) switchable polarization, (d) work of switching, (e) positive coercive voltage, and (f) negative coercive voltage. Note that a detailed definition of each parameter can be found elsewhere[2]. The solid black lines represent the same nanocapacitor in both of the PFM and BEPS images from Figures 1(c) and (d), respectively. The dark-brown colored region in Figures S3(e) and (f) represent the non-switched region. The scale bar represents 320 nm.

To further examine the local switching characteristics of each nanocapacitor, we observed BEPS spatial maps, which can be extracted from the phenomenological fitting of the PFM hysteresis loop at each grid point, as shown in Figure S3[3]. We observe that nanocapacitors generally show large positive (Figure S3(a)) and negative (Figure S3(b)) remnant piezoresponse characteristics even though there are some variations between nanocapacitors. Obviously, the work of switching and switchable polarization tends to be similar to remnant piezoresponse. This indicates that nanocapacitors show strong ferroelectric properties. Figures S3(e) and S3(f) represent spatial images of the positive and negative coercive voltages, respectively, which were obtained from the PFM hysteresis loop. Although there are some non-switched regions (dark-brown colored regions), the coercive voltages observed in the switched area are within the similar range regardless of the positive and negative voltages.

## 2. Current behavior of ferroelectric nanocapacitors

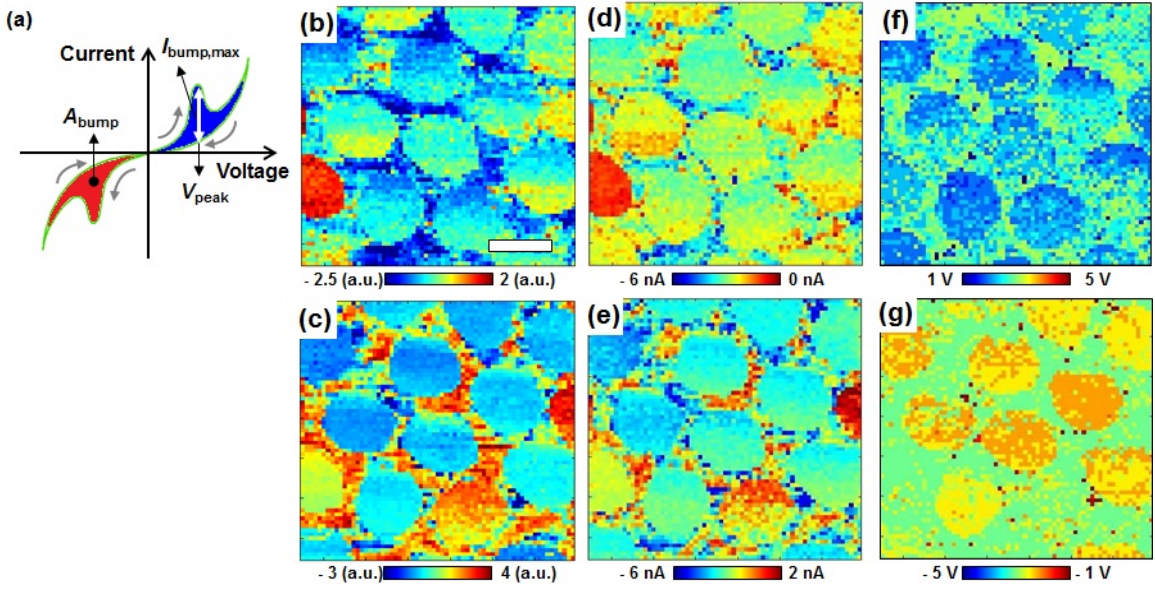

**Figure S4**. (a) Schematic of I-V curve and terms of each switching current component. (b, c) area of I-V curve, (d, e) maximum current of current bump, and (f, g) current peak voltage at (b, d, f) positive and (c, e, g) negative voltages in the ferroelectric nanocapacitors.

As shown in Figure S4, we analyzed the spatial I-V characteristics by observing spatial images of the area of the current bump ($A_{bump}$), the maximum current of the current bump ($I_{bump,max}$), and the current peak voltage ($V_{peak}$), as schematically presented in Figure S4(a). As shown in all of the spatial maps, most of the nanocapacitors show a uniform response within the nanocapacitor. The spatial maps of the area of current bump are similar to those of the maximum current of the current bump because the area is dependent on the intensity of the current peak. Figures S4(f) and S4(g) indicate that both current peak voltages show nearly the same contrast inside the nanocapacitors, although there are some variations between nanocapacitors. Interestingly, the current behavior is slightly different for each nanocapacitor. This can be explained by the fact that the ferroelectric switching is also different from each nanocapacitor, as previously reported[1] and shown in Figure S1. However, the observed spatial maps show nearly the same contrast inside the nanocapacitors, confirming that the I-V measurements were reliably performed within the nanocapacitors.

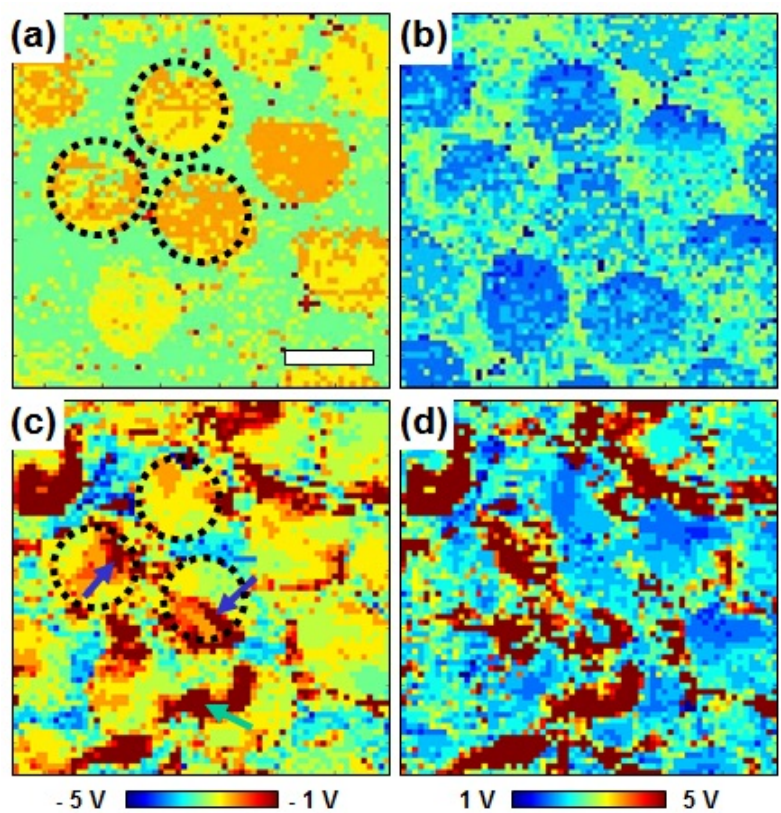

**Figure S5**. Spatial maps of (a,b) current peak voltages, *i.e., V$_{peak}$*, and (c,d) coercive voltages of PFM hysteresis loop for (a,c) negative and (b,d) positive biases, respectively, in the ferroelectric nanocapacitors. The current peak voltage was obtained from the I-V curve (Figure S4), and the coercive voltage was obtained from the phase change in the PFM hysteresis loop (Figure S3). The dark-brown colored regions in Figures S5(c) and (d)

represent non-switched regions. The color scales for the negative and positive spatial maps are the same. The scale bar is 320 nm.

In order to confirm that the current peak voltage, *i.e.*, $V_{peak}$, corresponds to the coercive voltage of polarization switching, we compared spatial maps of the current peak and coercive voltages, as seen in Figure S5, using the same color scale. Although the spatial map of the coercive voltage in Figure S5(c) shows local variations, the color of the peak voltage in Figure S5(a) is somewhat similar to that of the local coercive voltage inside the nanocapacitor (see the dotted line in Figures S5(a) and (c)). However, some of the pixels in the spatial map of the coercive voltage in Figures 5S(c) and (d) have a red color, indicating that the PFM hysteresis loop appears to be shifted toward the positive bias direction. In this region, the current peak voltage does not match with the coercive voltage. Nonetheless, most parts of the nanocapacitors show that the peak voltage is in accordance with the coercive voltage. This indicates that the current peak voltage is determined primarily by polarization switching, which generates mainly a switching current. Overall, the results obtained confirm that the current peaks in Figures 1(f) and 1(g) originate from the polarization switching current. Further studies may be necessary to clearly find quantitative spatial correlation between global switching current peak and local coercive voltage.

## 3. AFM-PUND results in ferroelectric nanocapacitors

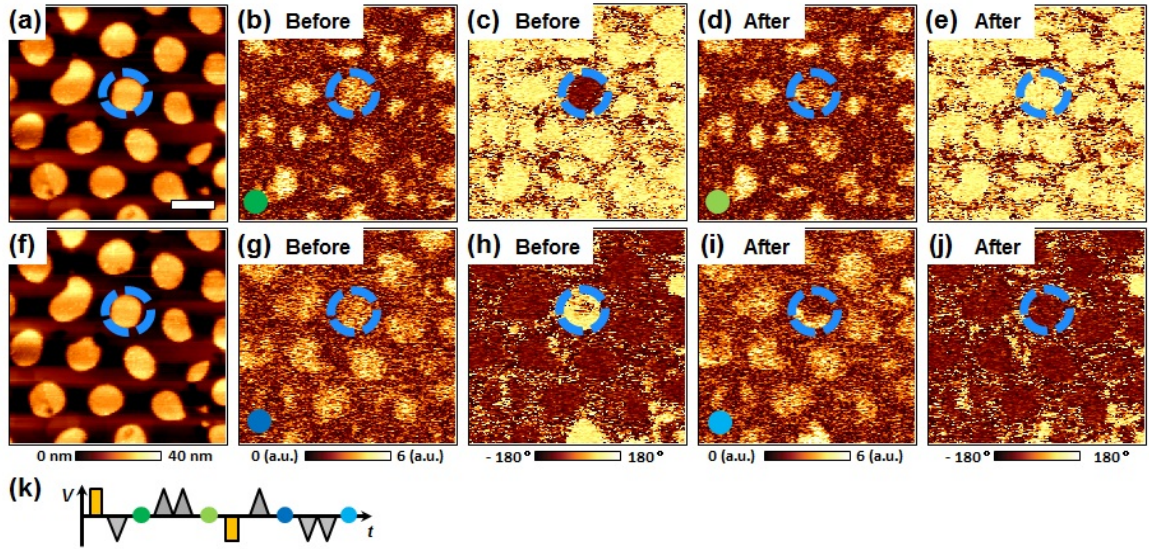

**Figure S6**. (a, f) Topography of the nanocapacitors, (b, d, g, i) PFM amplitude, and (c, e, h, j) PFM phase images, which were measured (b,c,g,h) before and (d,e,i,j) after application of (b-e) *PU* and (g-j) *ND* waveforms. Note that the maximum and minimum voltages of the AFM-PUND waveforms were +2.5 V (*PU*) and −2.5 V (*ND*), respectively. The scale bar is 400 nm. The blue dotted circles indicate a target nanocapacitor. Prior to the AFM-PUND measurement, to clearly show and confine the switched area within one nanocapacitor, background poling was first performed by applying voltages of (d) 3.5 V and (e) -3.5 V, respectively, over an entire area. Then, pre-poling was conducted by applying a unipolar triangular waveform to the single capacitor, down to a minimum of (d) -3 V and up to a maximum of (e) 2.5 V, respectively, as seen in Figures S6(d) and (e). (k) Schematic of the AFM-PUND waveform. The colored circles in Figures (b, d, g, i, and k) represent PFM imaging stages in the AFM-PUND sequence.

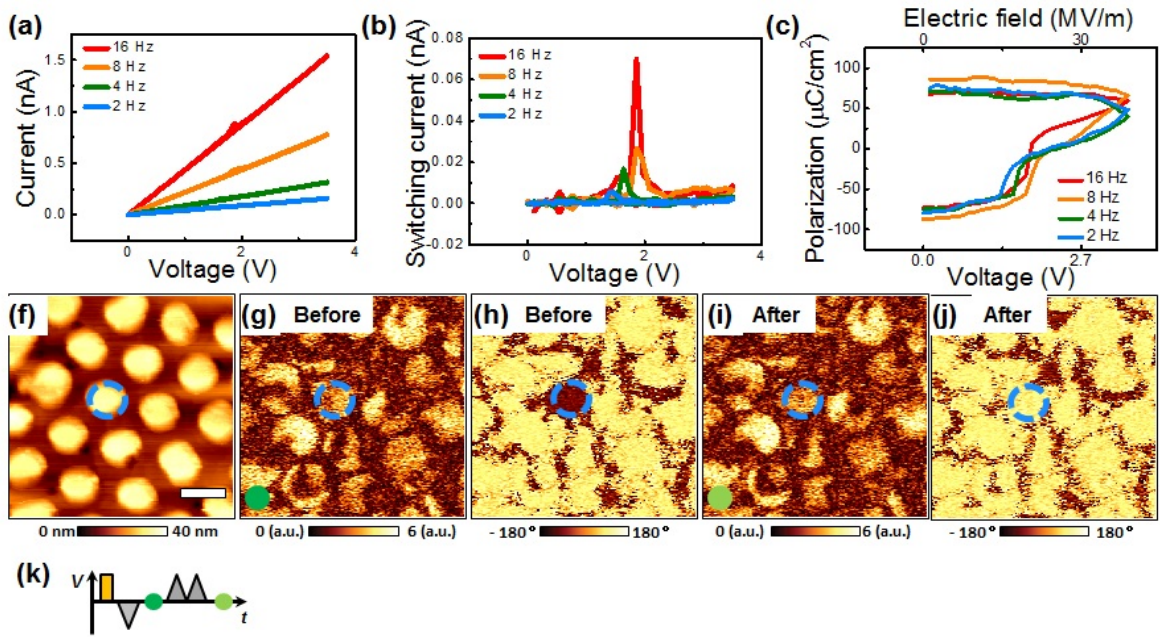

**Figure S7**. (a) IV curves acquired after *P* steps depending on the frequency of the applied *P* waveform in the ferroelectric nanocapacitors. (b) Switching current extracted from Figure S7(a). (c) P-E hysteresis loops for positive bias branches calculated from Figure S7(b). (f) Topography, (g, i) PFM amplitude, and (h, j) PFM phase images before (g, h) and after (i, j) application of *PU* waveforms with 2 Hz. The blue dotted circles indicate a target nanocapacitor. Note that the maximum voltage of the *PU* waveform is 3.5 V. The scale bar is 400 nm. Before the AFM-PUND measurement, background poling is performed by applying 3 V to the entire area. After that, pre-poling is performed by applying *N* unipolar waveform of - 2.5 V to the target nanocapacitor. (k) Schematic of the AFM-PUND waveform. The colored circles in Figures (b, d, g, i, and k) represent PFM imaging stages in the AFM-PUND sequence.

**4. AFM-PUND results in ferroelectric thin film**

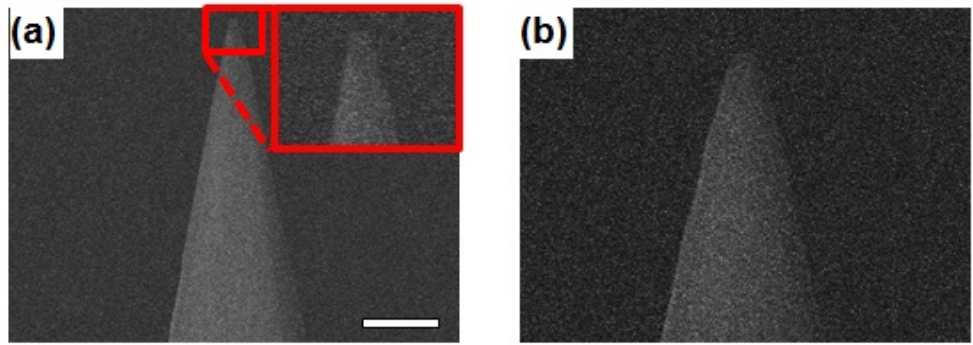

**Figure S8**. SEM images of a pure Pt tip: (a) pristine and (b) blunted states. The scale bar is 200 nm.

A pristine pure Pt tip with a radius of 15–20 nm (Figure S8(a)) was worn out by performing 25 consecutive scans with a set point of 700 nN. The radius of the blunted tip is around 60 nm. Even though the tip became blunted, its shape was maintained, as shown in Figure S8(b). Because the tip is a pure Pt tip and its shape is normal, it is believed that there is no serious electrical contact issue.

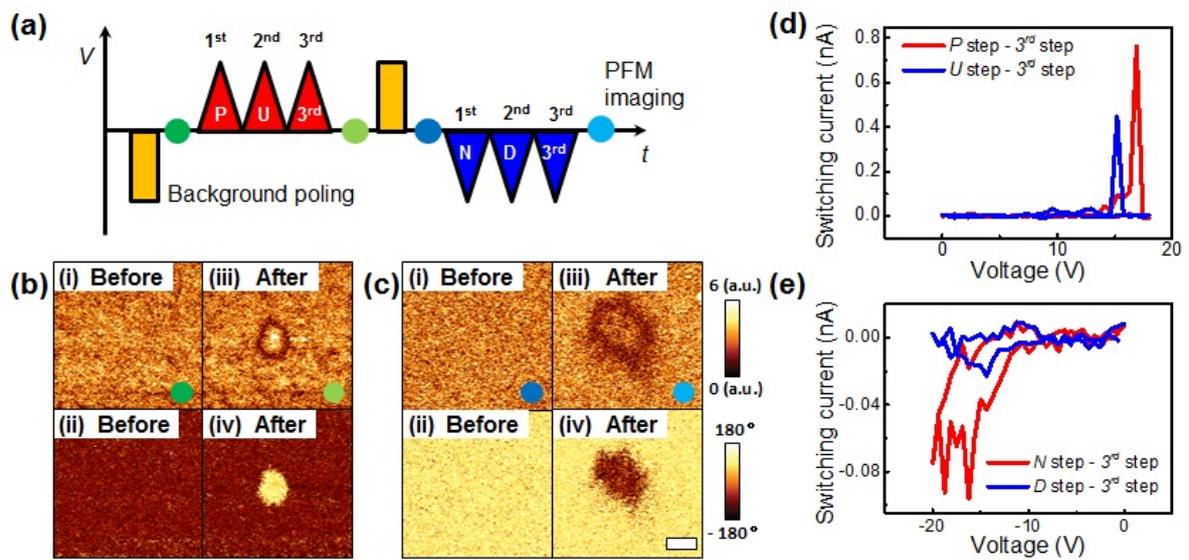

**Figure S9**. (a) Schematic of the modified AFM-PUND waveform in this measurement. The colored circles in Figures (a,b, and c) indicate the PFM imaging stages in the AFM-PUND sequence. (b,c) PFM (b (i, iii), c (i, iii)) amplitude, and (b (ii, iv), c (ii, iv)) phase images, which were measured (b (i, ii), c (i, ii)) before and (b (iii, iv), c (iii, iv)) after application of (b) *PU* and (c) *ND* waveforms, respectively, in the ferroelectric thin film. Maximum and minimum voltages of the AFM-PUND waveforms are +20 V (*PU*) and –20 V (*ND*), respectively. Figures (d) and (e) show the switching current, which was acquired by the modified (d) *PU* and (e) *ND* waveform, respectively. Before the application of the *PU* and *ND* waveform, background poling was performed by applying a voltage of (b) -20 V and (c) 20 V, respectively, over an entire area. Note that the 3$^{rd}$ step in Figures (d) and (e) represents the 3$^{rd}$ waveform shown in the modified AFM-PUND waveform of Figure (a). The scale bar is 400 μm.

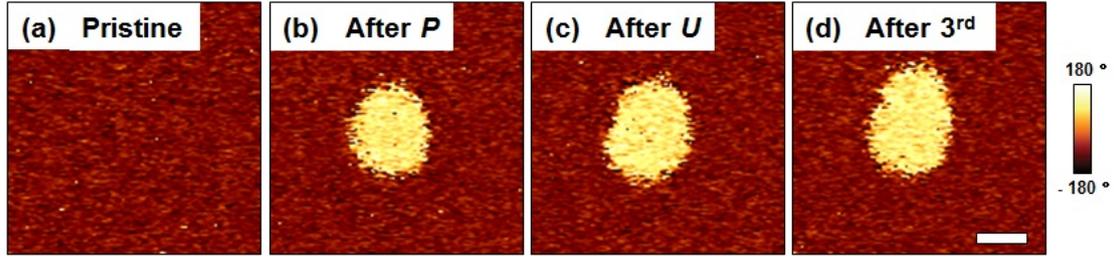

**Figure S10**. Phase images measured after (a) negative background poling, (b) *P* (c) *U* and (d) 3$^{rd}$ waveforms, respectively. Each switched area in Figs. (b, c, d) corresponds to 0.086, 0.103, and 0.110 μm$^2$, respectively. We note that these were obtained at different location with that in Fig. S9. Scale bar is 200 nm.

Figure S9 shows the switching current measurement without confinement of the switched area, i.e. without the pre-poling procedure, in the PZT thin film (see the difference between Figure 2(d) and Figure S9(a)). As shown in Figure S9(a), we applied a modified AFM-PUND waveform to measure the switching current. In principle, the switching current should not be generated in the second current measurement step (*U* and *D*) if there is no further switching. If the switched area can be confined within a specific area, i.e., the capacitor boundary, only one pre-poling procedure is sufficient to properly measure the polarization charge. However, if the switched area is not confined within a specific area, the switching current can even be measured at the *U* and *D* steps. In the case of the thin film, there is no capacitor boundary that acts as a barrier for the domain wall motion. Thus, whenever a voltage is applied to the sample, the switched area can be increased. Therefore, the switching current was even measured at the *U* and *D* steps (Figure S9(d) and S9(e)). Indeed, the increased switched area at the *U* step was observed in Figure S10. Further, the switched area should be confined in order to obtain accurate AFM-PUND measurements, and a background procedure is therefore required prior to the application of the AFM-PUND waveform for the thin film to confine the switched area.

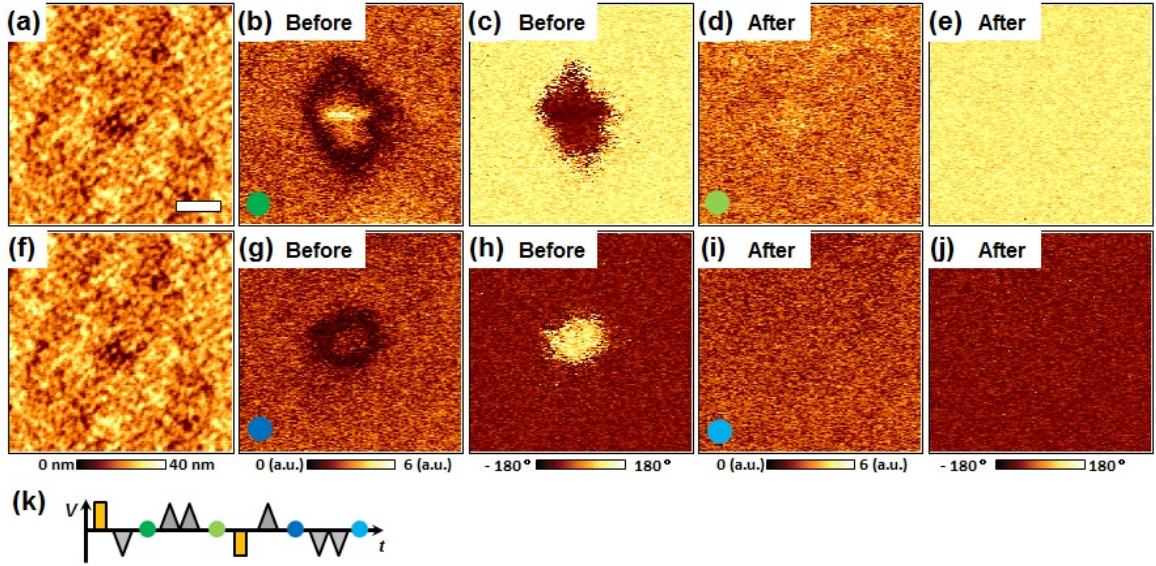

**Figure S11**. (a, f) Topography of the ferroelectric thin films, (b, d, g, i) PFM amplitude, and (c, e, h, j) PFM phase images (b,c,g,h) before and (d,e,i,k) after application of (b-e) *PU* and (g-j) *ND* waveforms. Note that the maximum and minimum voltages of AFM-PUND waveforms are +20 V (PU) and –20 V (ND), respectively. Similar to the measurement at the nanocapacitor, background poling was first performed by applying (a–e) 20 V and (f–j) -20 V, respectively, over an entire area, and then the pre-poling was conducted by applying a unipolar triangular waveform prior to the AFM-PUND measurement, down to a minimum of (b, c) - 20 V and up to a maximum of (g, h) + 20 V, respectively. (k) Schematic of the AFM-PUND waveform. The colored circles in Figures (b, d, g, i, and k) represent PFM imaging stages in the AFM-PUND sequence. The scale bar is 400 nm.

## 6. Large switching area in ferroelectric nanocapacitors

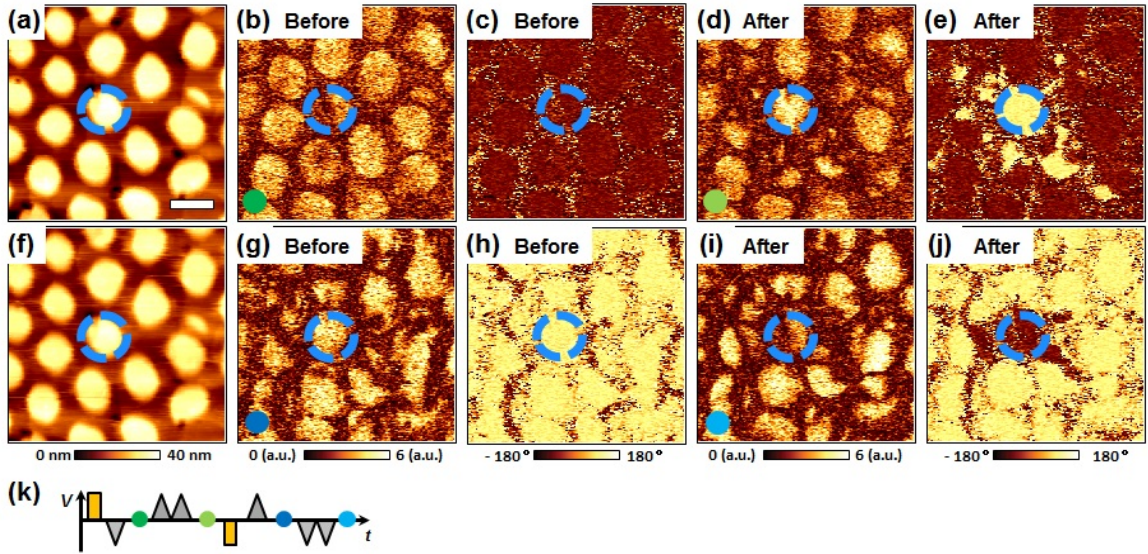

**Figure S12.** (a, f) Topography, (b, d, g, i) PFM amplitude, and (c, e, h, j) PFM phase images, which were measured (b,c,g,h) before and (d,e,i,j) after application of (b-e) *PU* and (g-j) *ND* waveforms, respectively, in the ferroelectric nanocapacitors. Note that the maximum and minimum voltages of the AFM-PUND waveforms are +4 V (*PU*) and -4 V (*ND*), respectively. (k) Schematic of the AFM-PUND waveform. The colored circles in Figures (b, d, g, i, and k) represent PFM imaging stages in the AFM-PUND sequence. The scale bar is 400 nm.

Generally, in the case of the capacitor, the switched area is confined by the capacitor boundary. However, in the case of the nanocapacitor, the switched area can be further extended beyond the boundary of capacitor if a relatively high voltage is applied to the capacitor[4]. To confirm this, we applied *PU* and *ND* waveforms with magnitudes up to a maximum of 4 V and down to a minimum of -4 V, respectively. As expected, the switched area is extended toward the outside of the target nanocapacitor.

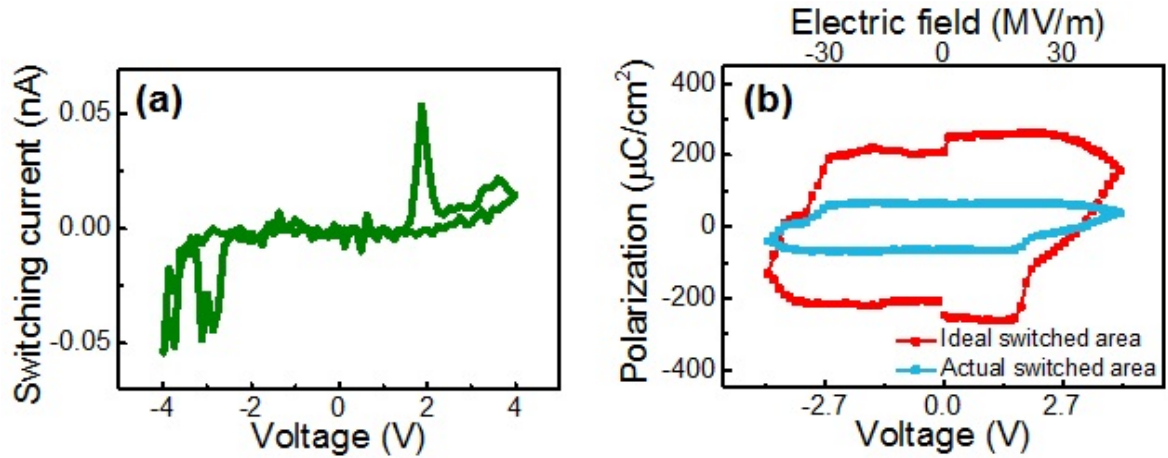

**Figure S13**. (a) Switching current measured in Figure S12 and corresponding (b) P-E hysteresis loop constructed from the switching current in Figure (a) in the ferroelectric nanocapacitors.

Based on the measured switching current in Figure S12, we calculated the polarization charge density to confirm whether it is within the acceptable range, despite having a large switching area beyond the capacitor boundary. The constructed P-E hysteresis loops are shown in Figure S13(b). When we consider only the top electrode area of the nanocapacitor as a switched area, the remnant polarization charge density was calculated as 200–250 $\mu$C/cm$^2$ (red colored curve in Figure S13(b)). However, this value is much higher than those obtained in previous studies. When the actual switched area measured by PFM was used to calculate the polarization charge density instead of the nanocapacitor area (see PFM images in Figure S12), it then falls within the reasonable range, i.e., around ±70 $\mu$C/cm$^2$. In other words, even if the actual switched area is much larger than the area of the top electrode, the calculated polarization charge density can be within the acceptable range based on the measurement of the actual switched area. However, in this case, note that the obtained polarization charge density ($\pm$70 $\mu$C/cm$^2$) may be slightly underestimated because $U$ and $D$ steps could induce a small portion of additional switching.